Article

# Multiscale Insights of Domain Unfolding in Fibrin Mechanical Response

Vivek Sharma[1] and Poulomi Sadhukhan[1,*]

[1]Department of Physics, Bennett University, TechZone II, Greater Noida, Uttar Pradesh - 201310, India.
[*]Correspondence: poulomi.sadhukhan@bennett.edu.in

ABSTRACT   Fibrinogen, the monomeric unit of fibrin, the main constituent of blood clot, has a very complex structure. The fibrinogen builds the fibrin fiber and network through the half-staggered packing via knob-hole interaction and the $\alpha$C crosslinkers. Due to its rich structure, the elastic behavior also shows a unique nature of very high stretchability and multiple regimes in stress-strain behaviour, which is not yet fully understood. We develop an Unfolding-incorporated Coarse-Grained Polymer (UCGP) model for fibrinogen to study the effect of domain unfolding on the mechanical behavior of fibrin fiber and network. Our model captures the stretching behavior of fibrinogen as observed in AFM and all-atom simulations. We further extend our model to fibrin fiber to study the effect of molecular unfolding at the fiber and network level. We anticipate that our model will be able to account for the nonlinear mechanical behavior of crosslinked fibrin gel. It is possibly the first model of this sort to consider the precise, controllable knowledge of the effects of domain unfolding in crosslinked proteins. This model can also be used to model systems that have sacrificial bonds.

## INTRODUCTION

Fibrinogen is a precursor of fibrin, which acts as the structural scaffold for blood clots. Fibrinogen is comprised of three pairs of polypeptide chains (A$\alpha$, B$\beta$, and a$\gamma$) that undergo many transformations when activated by thrombin, eventually resulting in a fibrin network. The main structural component of a clot is the fibrinogen, which is a fibrin monomer that is produced when thrombin cleaves fibrinopeptides A and B (1). These monomers polymerize and make a mesh-like structure that binds the blood cells, forming a blood clot. A molecular structure of human fibrinogen based on the X-ray structures of human fibrinogen and the fibrin fiber and D dimer as observed by Kollman et al. (2009) (2). In figure 1, a schematic diagram of fibrinogen, half-staggered configuration, fibrin fiber, and fibrin network is shown. Molecular structure of fibrinogen is shown in 1(A). After activation by thrombin, fibrinogen develop a pokect(hole) and a knob, which then leads to further polymerization by making double-stranded half-staggered configuration as shown by 1(B). 1(C) shows the fibrin fiber formed by further polymerization and parallel stacking of half-staggered configuration due to $\alpha$C connectors, and 1(D) shows fibrin gel network.

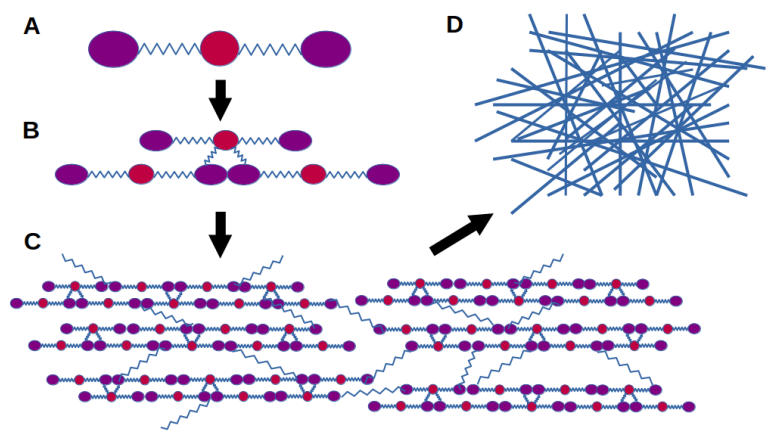

Figure 1: Schematic structure of (A) fibrinogen, (B) half-staggered packing of fibrinogen to form protofibril, (C) fibrin fiber formed by parallel stacking of protofibril due to $\alpha$C connectors, and (D) fibrin gel, where blue lines represent fibrin fibers.

Fibrin fibers have distinct mechanical properties critical for sustaining clot stability in the dynamic blood flow environment. These fibers are extremely extensible, expanding up to 2.5 times their original length, and have excellent elastic recovery (3–5). Furthermore, fibrin fibers undergo strain stiffening, making them more resilient to deformation as the strain increases, improving the mechanical stability of the clot (6). Fibrin fibers, under loading, change their structure and thus show nonlinear stress-strain behavior (7–9). The mechanical properties of fibrin fibers are determined by their molecular structure and interactions. Experimental and computational investigations reveal that the unfolding of particular protein domains contributes to the extensibility and elasticity of fibrin fibers. Unfolding of fibrinogen $\alpha C$ (9) and $\gamma$ nodule (10) contributes to fibrin fiber extensibility and elasticity. Unfolding of $\alpha$-helical coiled into $\beta$-sheets(11, 12) causes the strain-stiffening behavior of fibrin fiber. All-atom simulation shows the $\gamma$ nodule of the D-domains of the fibrinogen monomer gets unfolded from 5 nm to 80 nm as the pulling force is applied. This unfolding plays a crucial role in the extensibility and non-linearity.

Several attempts were made to get insights into the mechanical properties of fibrin fibers. People have used the worm-like chain model, but the worm-like chain model does not show the bilinearity as shown by the fibrin (9, 10, 13). A coarse-grained bead spring system with only three beads is used to model the fibrin network by S. Yesudasan et.al (14). In this work, the Heaviside step function is used with a cubic term, and the results are compared with AFM data, which is a good match in the lower region but deviates for high values of strain. N. Filla attempted to model the fibrin network by using a simple cubic and a body-centered mesh. He derives a bending potential that is used for calculating bending stiffness and persistent length, but does not include the effect of unfolding. (15). F. Maksudov gives a fluctuating bilinear spring model and tries to explain the rapture behaviour of the fibrin fiber and shows the bilinear behavior (13).

Despite breakthroughs in the research of fibrin structure and mechanics, it is still challenging to completely comprehend the link between fibrinogen's molecular structure and the macroscopic behavior of fibrin networks. Existing models do not fully describe the role of domain-unfolding in fibrin fibers' bilinear mechanical response, which consists of an initial soft phase followed by a strain-stiffening phase and the nonlinear elasticity found at the network level. Multiple factors influence these occurrences, including fiber alignment, cross-linking, fibrin mesh hierarchy, and unfolding of the globular domains. More research is needed to determine the roles of these components in the overall mechanical behavior of fibrin in healthy and pathological circumstances. Several models exist that try to explain the mechanical behavior of the network. Also, an all-atom simulation study for a single fibrinogen monomer is present, which shows unfolding and microscopic details(16). But there is no bridge that links the smaller unit to the actual large network. It is very tedious and computationally nearly impossible to take all-atom simulations to the network level. So to fill this monomer-to-network gap, we need a model that can be taken up to the network level while taking care of unfolding, which is happening at the microscopic level.

In this article, we focus on the effect of unfolding, particularly on the $\gamma$-nodule unfolding, as it can play a vital role in the extensibility and the force-extension behavior of the fibrinogen and fibrin fibers. We develop an unfolding-incorporated coarse-grained polymer (UCGP) model for a fibrinogen monomer that incorporates the $\gamma$-nodule unfolding. Using a bottom-up approach with this single monomer, we have built a fiber and have used that fiber to build a network of fibrin fibers that includes domain unfolding. We use molecular dynamics (MD) simulation to study the dynamical behavior of the coarse-grained model of fibrinogen and fibrin fibers.

## MODEL AND METHOD
### i. Fibrinogen

Fibrinogen has three globular domains (D-E-D) connected with three $\alpha$-helical coils that can be seen by high-resolution atomic force microscopy (17). Each of the D and E domains is represented with three beads connected via two springs, and each of the two coiled parts is represented by two springs as shown in figure 2, totaling 11 beads to represent the whole fibrinogen(18). In figure 2a, the upper picture is an all-atom simulation taken from the work by Zhmurov *et al* (16), and the bead-spring system used for the coarse-grained model is shown below.

The length of fibrinogen is taken as 45.5 nm and the diameter of a single fibrinogen is taken as 4 nm, nearly equal to the reported value in Ref. (16, 19). A fibrinogen has one E-domain of 1.5 nm, connected to two D-domains of 5 nm each via coiled-coil structures of length 17 nm on each side. The angle between bonds is initially taken as 180°, which is slightly greater than as suggested by Yermolenko (17). Mass distribution is taken as S. Yesudasan et.al (20).

Different spring stiffnesses have been chosen for the springs used for different domains. We denote $k_C$, $k_D$, and $k_E$ as the spring constants for the coiled-coil region, the D domain, and the E domain, respectively. The spring constants are estimated assuming Hookean behavior of the system to mimic the overall force-extension behavior as observed in all-atom simulation (16). We choose the rest lengths of the springs $k_D$, $k_C$, and $k_E$ are 2.5, 8.5, and 0.75 nm, respectively.

The unfolding of the D-domain happens in three steps. To incorporate this, we have changed the spring parameters three times. First, the stiffness and the rest length are both changed. For the second and third steps, only the rest lengths are changed.

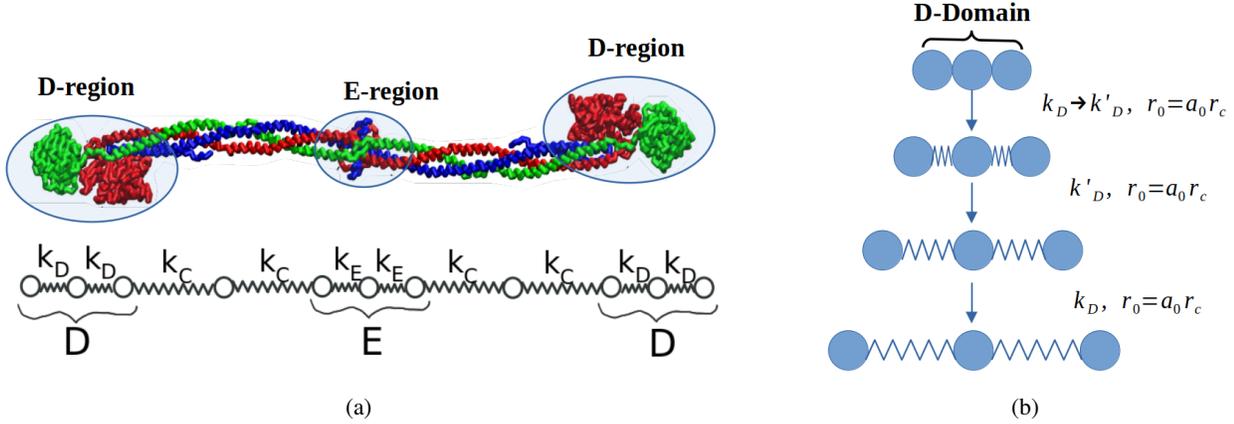

Figure 2: (Left, Fig. 2a) Schematic diagram of fibrinogen molecule taken from A. Zhmurove et.al (16). $k_D$, $k_C$ and $k_E$ denote the spring constants of D, C, and E regions. (Right, Fig. 2b) The three-step unfolding criteria of the D domain.

After complete unfolding of the monomer, the D-domain changes its length from 5 to 80 nm, the E-domain changes its length from 1.5 to 9 nm, and the coiled region goes from 17 to 41 nm. Now, if we calculate the ratio for $k_C:k_D:k_E$, that comes out to be 1:3:10. The chosen spring constant values are displayed in Table 1. The condition is given on stress values such that if the stress value in the D-domain is higher then 400 in LJ units then first unfolding takes place and the spring parameters are changed according to $r_0 \to r_0^{(1)} = a_0 r_c$ where $r_c$ is the current length of the spring and factor $a_0$ can take values form 0.6-1.0 (18). The value of $k_D$ changes to $k'_D$, which is 10 pN/nm for this case. For the second and third unfolding, only the values of the rest-lengths of the springs are changed at higher values of stress according to $r_0 \to r_0^{(i)} = a_0 r_c (i = 2, 3)$. Figure 2b depicts the step-wise criteria of unfolding. Table 1 summarizes all the spring parameters of each domain.

| Bonds | Stiffness(pN/nm) | Rest length(nm) |
|---|---|---|
| D-Domain | $33 \to 10$ | 2.5 |
| E-Domain | 110 | 0.75 |
| Coiled region | 11 | 8.5 |
| Knob-hole | 300 | 0.75 |
| $\alpha$C | 15 | 15 |
| D-D | 300 | 4 |

Table 1: The details of spring connectors of each region, spring stiffness, and rest lengths of the springs.

We tune the spring parameters and the threshold stress values of stress at which the unfolding occurs, such that our model can reproduce the observed behavior in all-atom simulations. We get six peaks due to six unfolding, three on each side of the monomer, with the overall slope of the $f$-$x$ graph unity. Due to unfolding, there should be a drop in the force value, and after every drop, the force again starts to build up as we stretch the monomer further.

## ii. Protofibril

Fibrin protofibril is a half-staggered arrangement of fibrinogen, which is the first step of polymerization towards the construction of a fibrous structure (21–23). The protofibril is formed by the "knob-hole" interaction, which plays a key role in the stability and strength of fibrin fibers (23). The knob-hole interaction is a strong short-range interaction and also has a large affinity (10, 24).

So we have taken high spring stiffness ($k_{kh}$ = 300 pN/nm) for this bond and small rest length, $r_{kh}$ = 4 nm. That is approximately equal to the diameter of the monomer. A schematic diagram is shown in Figure 3. An interesting fact is that during the formation of fibrin fibers through knob-hole interaction, the D domains come even closer (25) and interact very strongly, so these interactions are modeled using a spring with a very short equilibrium length of 0.75 nm and high stiffness value ($k_d$ = 300 pN/nm).

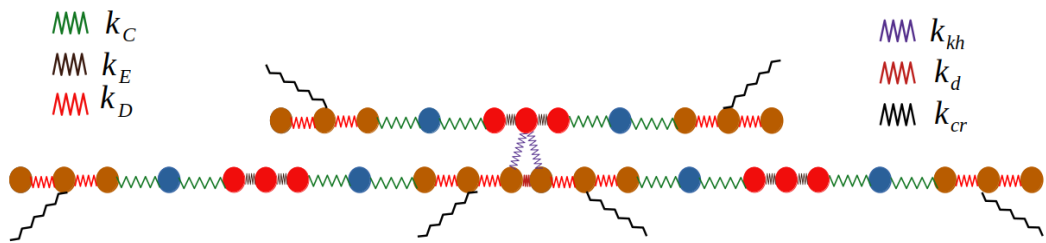

Figure 3: Schematic diagram for a fibrin protofibril to model fibrin fibers. Different colors show different domains and the springs connecting those. $k_{kh}$, $k_d$ and $k_{cr}$ are the additional spring attached to the model. The $k_{kh}$, $k_d$ and $k_{cr}$ represents the spring constants for knob-hole, D-D and $\alpha$C crosslinkers respectively.

### iii. Fiber

In the formation of fibrin fibers and networks, an additional important interaction is crosslinking via $\alpha$C (26–28). Crosslinking is important for parallel stacking and the varying thickness of the fiber (29). We have built a fiber by a parallel arrangement of protofibrils connected by $\alpha$C crosslinkers. Figure(4) shows the schematic diagram of the arrangement of the protofibril into a fiber. The number of protofibril strands is related to the thickness of the fiber. In fibrinogen, $\alpha$C crosslinker chain is found to be 21 ± 6 nm, which is much smaller than its contour length estimated by the total number of residues present in $\alpha$C chain (30). Upon stretching the crosslinkers, it gradually opens up to full length. Hence, we consider the crosslinkers as a very soft bond with a small spring stiffness, $k_c$ = 15 pN/nm. The rest-length is chosen approximately the same as the initial distance of the beads connected by crosslinkers, $r_c$ = 15 nm. In figure 3, each interaction is represented by a different kind of springs having a different color. In our model, we have control over each interaction by changing the spring parameters that represent that particular interaction. To study the effect of unfolding in a fiber while stretching, we present here three different sizes of the system, as summarized in Table 2, and the structure is depicted in Fig. 4. System1 has two protofibrils in parallel and 10 monomers of fibrinogen in horizontal and only 40 monomers in total. For System2, we took 5 protofibrils in parallel, keeping the length as of System1, thus it has 100 monomers. In System3, we increased both length and diameter. It has 15 parallel protofibrils and 30 monomers along the length. This system has 900 monomers. In this manner, we can design different sizes of systems, step by step, moving towards larger systems and finally to a fibrin network.

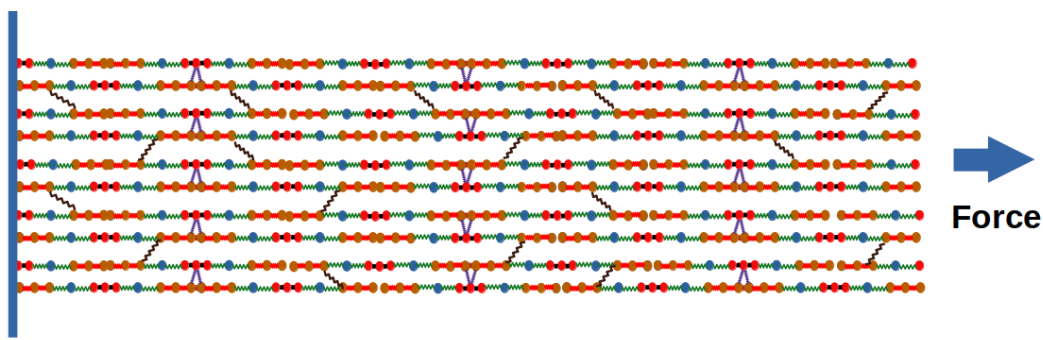

Figure 4: Schematic diagram for a fibrin fiber as used in our model. A parallel arrangement of protofibril strands interconnected by $\alpha$C crosslinkers (black springs). The left end of the fiber is anchored, and a stretching force is applied to the right end.

| Name | System1 | System2 | System3 |
|---|---|---|---|
| Dimension | 2×10 | 5×10 | 15×30 |

Table 2: The dimensions of fibers taken for simulation. $n \times m$ represents the fiber dimension where $n$ is the number of protofibrils taken in parallel and $m$ represents the total number of monomers taken lengthwise.

## iv. Network

A fibrin network is made up of fibrin fiber bundles stacked with each other because of cross-linking. For the formation of a network, it is necessary to allow branching. So simple parallel arrangement of fibers won't work here. In general, the polymer mesh is modeled as a triangulated network. We follow the same structure and restrict to two dimensions only, as stretching behavior won't be affected significantly by the transverse direction. At this point, to scale up the structure, the model is coarse-grained further. Here we take a single spring behaving effectively as a fiber segment undergoing unfolding. To do that, we have followed the protocol mentioned below.

(i) A square area is filled with points by using the Poisson disk method. The radius of the Poisson disk is taken as 5 nm, and the height and width of the box as 100 nm. A triangulated network is constructed by the Delaunay triangulation method. The network is diluted by deleting some of the bonds randomly, keeping at least three bonds per node to maintain the integrity of the structure even at high strain (Figure 5).

(ii) In this mesh, each connection between the nodes represents a segment of the largest fibrin fiber we have modeled, through a single spring. The equilibrium length of the spring is taken as 5 nm, and the stiffness is taken as 3 pN/nm before unfolding and 3.75 pN/nm after unfolding.

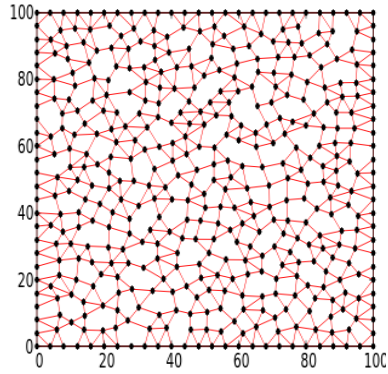

Figure 5: Triangulated network of fibrin fiber.

For the network also a similar scheme of pulling is used as is used for fibers. One boundary of the square is kept fixed, and the other end is moved.

To incorporate unfolding at the fiber level, we have used the condition that if the stress value in the D-domain is greater than a specific value, then the parameters are changed. The same can also be used for the network, where if the stress value at a node reaches a specific value, then unfolding starts. But in a network, one node is connected to multiple fibers. The stress value at a node may be high due to any of these connections, so if we use the same condition, then it is possible that out of three connections, stress is rising due to only one connection, but due to the condition, all three bonds get unfolded. So we have used a little modified strategy to be able to analyze larger systems more effectively. To deal with this problem, we must go by the stress created on individual bonds, which depends on the bond lengths. The bonds that have maximum extension have higher stress value and should be unfolded, while the other bonds at the node should not change if there is not much extension in those bonds. So, for the network, we use the condition on the bond-length that if the bond-length of any bond becomes larger than a particular value, then the parameters of the springs will change. Moreover, at the network level, there are multiple components through which the stress can be released; the amount of unfolding should reduce at every step. Therefore, we have used an additional condition that after every time step, only 10% of the eligible bonds will undergo unfolding.

In figure 5, a triangulated network is shown, which is used for the simulation of the fibrin network. In this network, each fiber is denoted by a simple spring having a rest length of 5 nm. Each connection resembles a System3 fiber, as if the length of the fiber is scaled to 5 nm. From our model, we find the initial stiffness for fibrin 3 pN/nm, and after unfolding, the overall stiffness is 3.75 pN/nm. These same values are used for the connecting springs used in the network. If we measure the change in length due to the unfolding in the System3 fiber, we find that the change in length is about $1/4^{th}$ of the original length taken. The same is used for the network, the unfolding occurs when the extension of a spring is 1.25 nm, which is $1/4^{th}$ of the original length of a segment.

## MD Simulation Details

We use LAMMPS (Large-scale Atomic/Molecular Massively Parallel Simulator) for the classical molecular dynamics simulation (31). The simulator uses Langevin dynamics to evolve the system. The temperature has been kept fixed at 300K. For the bonded interaction via the springs, the harmonic potential $k(r - r_0)^2$ is used. For the non-bonded interaction, `soft` potential is used, where the interaction energy is given by $E = A \left[ 1 + \cos \frac{\pi r}{r_c} \right]$ for $r < r_c$. It is used to push apart the overlapping atoms for $r < r_c$. To provide bending stiffness to the polymer, the harmonic angular interaction, $k_\theta (\theta - 180°)^2$, is used. After initial relaxation steps, a constant force is applied at one end by `fix move linear`, and force-extension data are noted. Step size $dt$ for the simulation is taken as 0.005. For `soft` potential, the cutoff values $r_c$ are taken to be 0.7 nm for E-domain and 2 nm for all other atoms, and for $A$, the value is taken as 0.4 pN.nm.

## RESULTS & DISCUSSION

### i. Fibrinogen

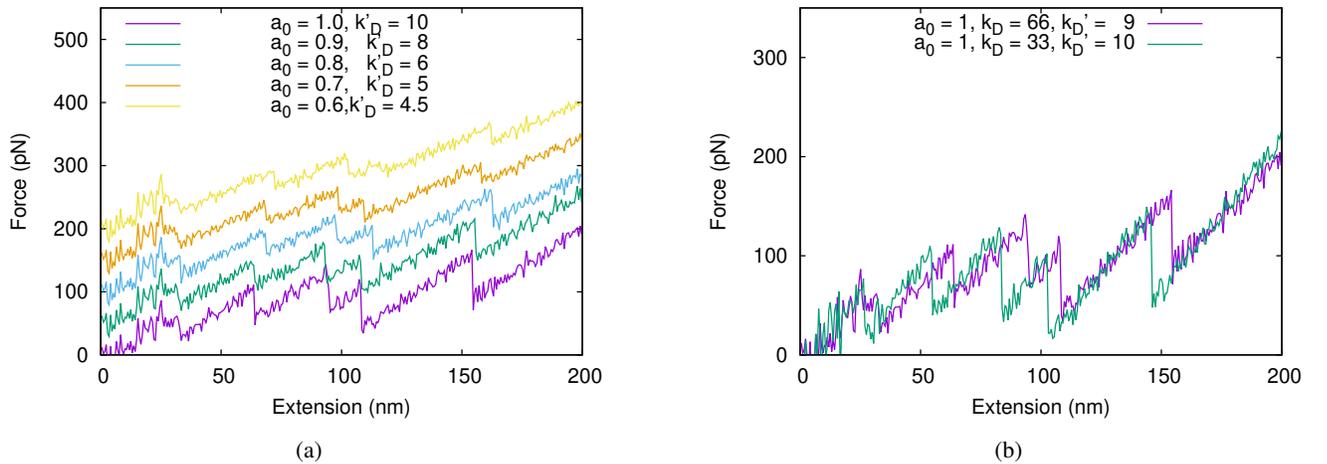

Figure 6: Force-extension curve obtained from the simulation of fibrinogen stretching. Fig. 6a shows how the behavior changes with the choice of $a_0$. For different values of $a_0$, $k'_D$ is adjusted to get the correct slope and peak position with $k_D = 33$ pN/nm. Fig. 6b shows how the force-extension behavior changes for different values of spring stiffnesses, $k_D = 33$ pN/nm and $k_D = 66$ pN/nm, with $a_0 = 1$.

To study the force-extension behavior, we keep one end of the system fixed, and the last beads at the other end of all strands are pulled by a force. After every step of pulling the system, a certain number of relaxation steps are run to rearrange the beads so that the stress gets distributed to other parts of the system. Once the unfolding criteria are met, the D-domain starts unfolding, and the stress on the atoms gets relieved. Figure 6 shows the unfolding of two $\gamma$-domains in three steps with six peaks (18) (more precisely, six dips). The first two peaks are very small and very close, while the other peaks are clearly visible. We have plotted different combinations of $a_0$ and $k_D$. To avoid overlapping graphs, we shift the scale for each curve by 50 units for better visibility. We observe in Fig. 6b that for a combination of $a_0$ and $k'_D$, we get the same behavior with the same number of peaks and the same slope. This provides us with a range from which these two parameters can be selected. If we change the stiffness of the springs while keeping the ratio the same as stated in Model and Method, Section (i), we also have a similar behavior with a little change in the position of the peaks. As long as the ratio of the stiffness for all three regions is kept fixed, the initial value chosen for $k_D$ doesn't affect the graph much.

This step of fibrinogen stretching can be considered as the calibration and validation step for our model to reproduce AFM stretching and all-atom simulation results, as shown in Ref. (16). We keep the parameter unaltered as we scale up the system to construct protofibril and fiber, and compare our simulation results with the experimental data.

### ii. Protofibril

We have taken a double-stranded protofibril that has 10 monomers in each strand; thus, a total of 20 monomers in a single protofibril. It has a diameter of 8 nm and a contour length of 455 nm. After equilibrating the system without force, it achieves an end-to-end length of 340 − 346 nm due to the entropic effect. The extension is measured with respect to this relaxed length.

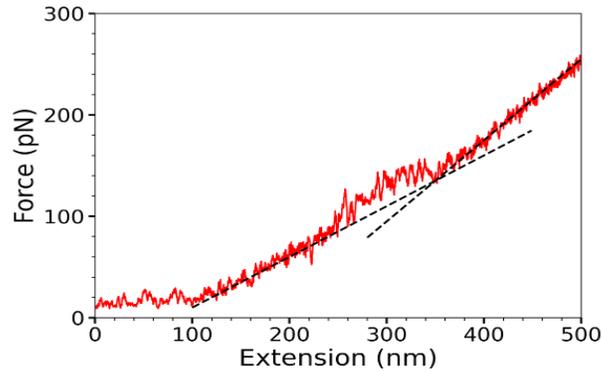

Figure 7: Force-extension graph of a protofibril having a length of 10 monomers. The black dashed line guides the slope change before and after unfolding.

Figure 7 shows the result of pulling a double-stranded protofibril. We see that the initial slope is very low as the structure is getting stretched at a small force. As the system reaches its contour length, the stress starts building up in the bonds, resulting in a change in slope. It may be noted that this transition happens at an extension of about 110 nm, which is approximately the difference between the rest length and contour length, where the enthalpic regime sets in. Upon further increase of force, the unfolding of $\gamma$-nodule starts, and all unfolding gets completed by 400 nm. After that, the force-extension behavior is linear with a slope of 0.75 pN/nm, which is obvious for a series and parallel arrangement of Hookean springs.

The existing bead-spring models of fiber stretching (14, 20) without having domain-unfolding in the model, show the entropic to enthalpic transition in the force-extension graph through a change of slope. Once the enthalpic region is reached, it should show the same slope afterwards for the Hookean spring model. If the spring has a cubic term or WLC-like terms, then we see a nonlinear increase in the force-extension curve. As we have unfolding in our model, we see the effect of unfolding in the enthalpic region. It changes the continuity of the slope in the region where unfolding continues to happen. In the next section, we shall see that it becomes more prominent as the size of the system increases.

## iii. Fiber

To study the behavior of fiber stretching, we have studied both the cases of crosslinked and uncrosslinked fibers.

### iii-a. Uncrosslinked Fiber

In this case, we have taken the scenario where there are no $\alpha$C connectors. This can be achieved by removing those or prohibiting the crosslinking (32).

The average force was calculated for parallel stacking by dividing the total applied force by the number of protofibril strands incorporated into the system. As a result, the system behaves as if the force is applied individually to each protofibril strand. To ensure comparability between systems and isolate the effect of unfolding, force averaging is performed, wherein the total applied force is normalized by the number of parallel strands used in constructing multistrand fibers.

In figure 8a, we plot force vs. extension of System1 without any crosslinking. There are four regimes in the plot. The first two are the expected entropic and enthalpic regimes, then a plateau-like region in the range 120 − 450 nm where most of the unfolding takes place. The fourth region is again a linear increase of extension with force with a slightly higher slope, 0.75 pN. In the third region, the stress builds up enough in the fiber, and the unfolding of the $\gamma$ nodule starts. As we have seen in the case of fibrinogen, that unfolding reduces the force on the system at constant extension, naturally, the slope decreases on the $f$-$x$ curve in the third region where the unfolding occurs. The bottom row in Fig. 8 displays the number of unfolding with the applied force value. Surprisingly, we find that the unfolding does not occur in a continuous manner; first unfolding happens when a system moves from the entropic region to the enthalpic region. It shows a double-peak distribution for all the fiber sizes and for both crosslinked and uncrosslinked fibers. The distribution becomes sharper and more localized in force as the system size increases. For a smaller size, the unfolding starts early and is more scattered in force due to more fluctuations.

Figure 8b shows the force-extension behavior of system2. We observe similar behavior to that of a single protofibril. Typically, a fibrin fiber has a diameter of about 100 nm (32). System3 is 120 nm in diameter and 1.4 $\mu$m in length. The plot for this system is shown in fig8c. The unfolding completes at 1050 nm, which is a similar value to that in the smaller system when we scale the length of the system. Here we observe that as the size of the system increases, the fluctuations decrease in $f$-$x$ behavior and unfolding pattern, but the overall features remain the same.

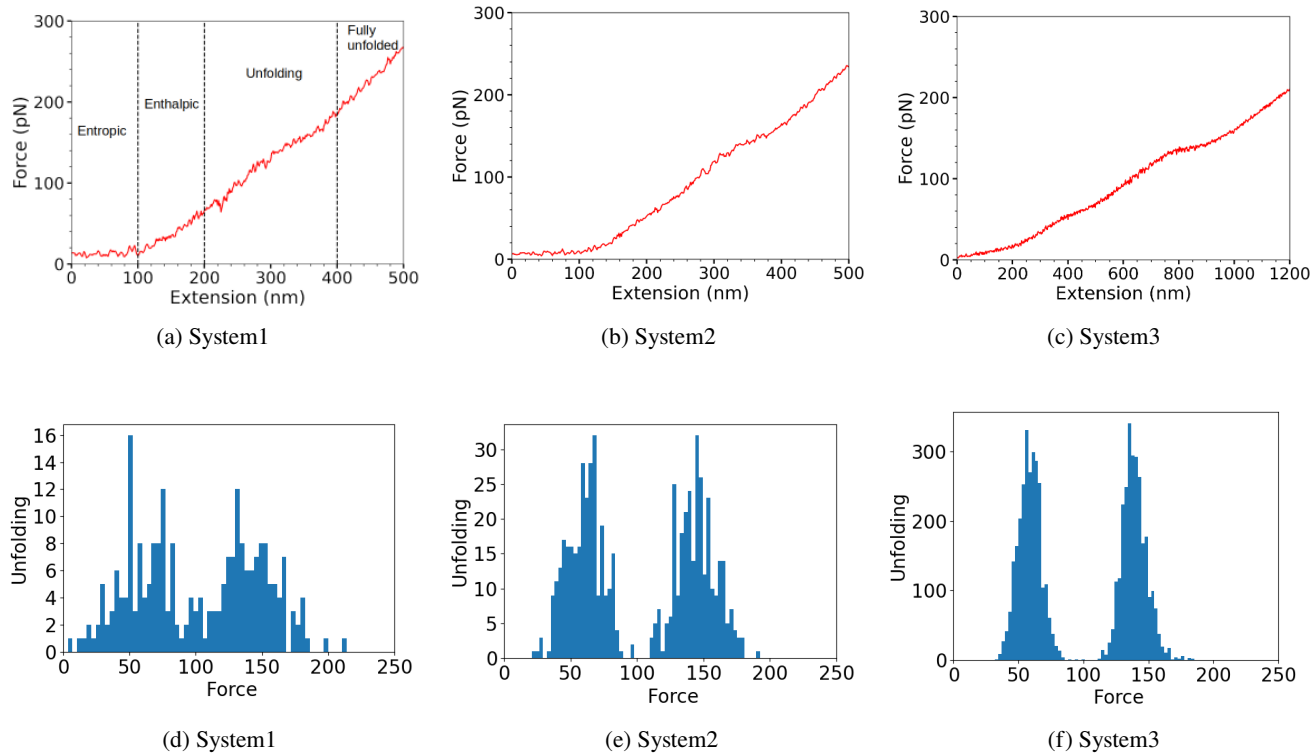

Figure 8: Top row (a,b,c): The force-extension behavior of System1 (2 × 10), System2 (5 × 10), and System3 (15 × 30) without any crosslinking. Here $n \times m$ represents the dimensions of the fiber, $n$ representing the number of protofibrils taken in parallel, which determines diameter, and $m$ represents the number of monomers taken in series, which determines length. Bottom row (d,e,f): The unfolding pattern, the number of unfolding against force is plotted for Systems1,2, and 3, respectively.

**iii-b. Crosslinked Fiber**

In crosslinked fibers, $\alpha$C crosslinkers are connected from one protofibril's D-domain to other ones. We have varied the crosslinker concentrations and studied the effect of domain unfolding in the force-extension behavior. The concentration is calculated as the number of crosslinkers present divided by the total number of $\alpha$C binding sites. So if all the $\alpha$C are connected, then it is taken as fully crosslinked, or 100%, while if 50% of those are connected, then the concentration is said to be 50%, and so on.

In Figure 9, we see force-extension for all the systems with 15% $\alpha$-C crosslinking; Figure 9a is for System1, Figure 9b is for System2, and 9c for System3. We see the unfolding pattern is almost the same for 9a and 9b. The first unfolding takes place when the system moves from the entropic region to the enthalpic region. After that, it incorporates the changes due to unfolding, and then, after some time, when the system reaches almost 75% of its original length, further unfolding takes place. Here, the pattern and behavior are almost the same as for the case of no crosslinking. The only change we see is the overall change in slope for all of the systems. The slope gets increased due to the crosslinking, which is expected as a new binding interaction is added to the system.

From the unfolding pattern for the crosslinked case (bottom row in Figure 9), we find that the first peak is larger compared to the second one. This becomes even more prominent for a larger size of the fiber. This happens as the fibers become more connected via crosslinkers; the force applied at one end, more evenly affects all domains, thus more domains unfold at one go.

Figure 10, the force-extension data of a fixed-sized fiber, shows the effect of unfolding as the concentration of crosslinking is increased. It is displayed only for System2 here; for other sizes also, the main observations remain the same. The effect due to increased crosslinking was suggested in (32). They have discussed that with the increase in crosslinking, the stiffness of the fiber increases. Here, one can observe that as the crosslinking increases, that is, as we go from the case of uncrosslinked to the fully-crosslinked, the slope of the overall graph increases from 0.75 to 0.95 pN/nm. When the crosslink concentration is

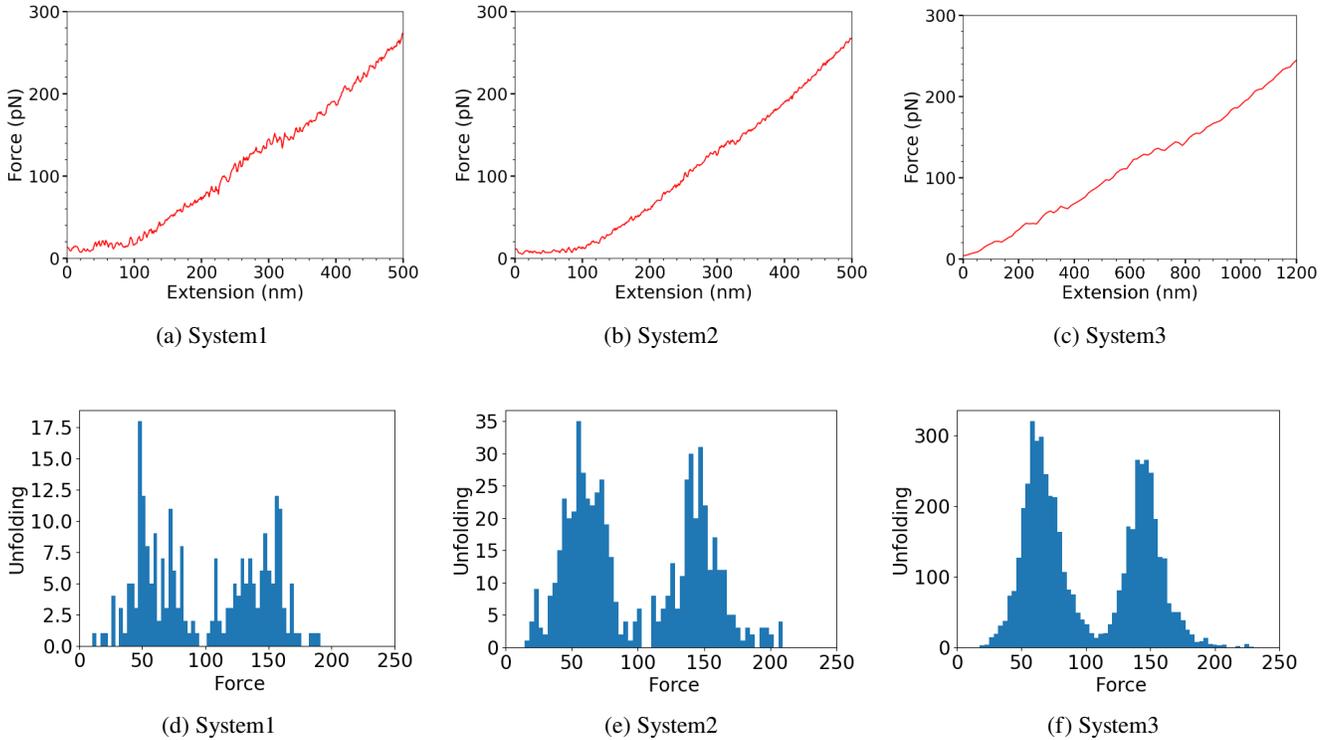

Figure 9: Top row: The force-extension behavior for crosslinked fibers, System1 ($2 \times 10$), System2 ($5 \times 10$) and System3 ($15 \times 30$), respectively. Bottom row: The unfolding pattern for System 1,2 and 3 respectively.

100%, the effect of unfolding is almost negligible. It is very much predictable because, as the number of crosslinkers increases, the bundle is more tightly packed and unfolding is hindered. For a lower concentration of crosslinkers, we see the effect of unfolding prominently. As concentration increases, the effect of unfolding becomes more prominent in an earlier stage when the system moves from the entropic to the enthalpic region, as shown in Fig. 11. The first peak is higher than the second one, which indicates that as the concentration of $\alpha$-C crosslinkers increases, all the unfolding happens instantly as the stress starts to build up in the molecule. It is to be noted here that the plateau-like region, which appears due to the unfolding of domains, appears within 100% of strain for all system sizes and concentrations of crosslinkers.

## iv. Network

As outlined in the Model and Method section, different strategies were employed for network deformation. Here, we present cases in which all bonds undergo unfolding when a node experiences stress: (i) over three steps and (ii) within a single time step. In these scenarios, all stressed bonds unfold without restricting the unfolding process to 10% of the total bonds.

Figure 12a shows the results of stretching the fibrin network with three-step unfolding, as is used in the case of fiber. For this case, we have no control over which bond is responsible for stress at the node. It follows the simple rule that if the stress condition is reached, all bonds at the node will change. One more important thing is that in this case, the rest length is changed with the current length at every unfolding. Therefore, in a way, this case presents the maximum effect of unfolding that we can see in a traingulated network. It shows a wide plateau-like region.

Figure 12b shows the force-extension behavior of the network with effective one-step unfolding, which captures the net change in length of three-step unfolding. Here the rest is changed from 5 nm to 10 nm after unfolding, keeping the spring stiffness unaltered. Here also all the stressed domains are unfolded. We can see a prominent region with a lesser slope where the unfolding of domains happens.

This is remarkable result that for a fibrin network, we see a now four regions, viz. (i) initial entropic region with a very low slope, (ii) enthalpic region representing stiffening of the network and fibers, (iii) a region with low slope where the built-up stress on fibers get released through unfolding of domains, and (iv) a region with high slope represeting a fully unfolded network. This region will show a strain-stiffening, and the slope can increase even in a steeper manner if the stiffness after unfolding is increased in the domains. Similar behavior is observed in the fibrin gel shearing experiment, where differential modulus $K'$ shows multiple regimes with a plateau region when plotted against stress $\sigma$ (9). Therefore, our UCGP model can give an important hint on the possible mechanism that may be present during gel shearing and stretching.

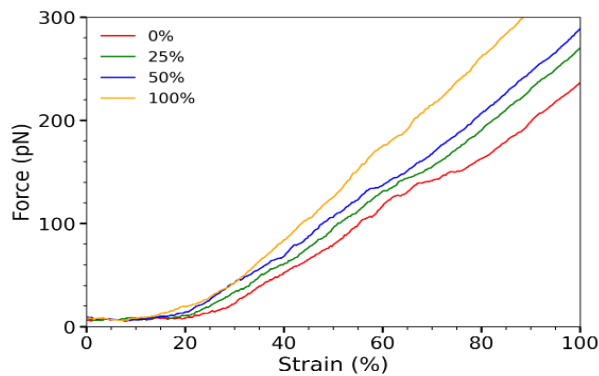

Figure 10: Force-extension relation for different concentrations of crosslinkers in System2 (5 × 10 fiber).

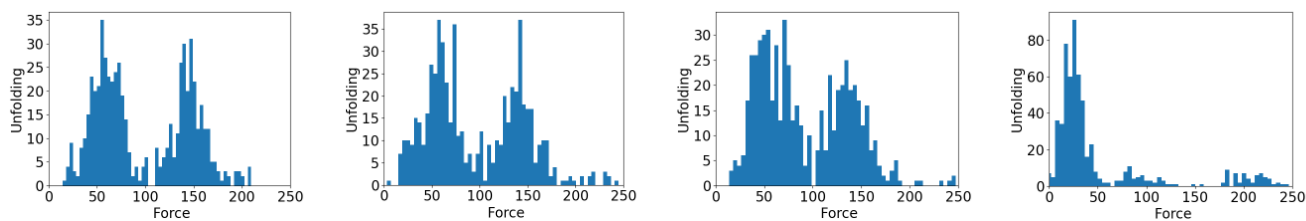

Figure 11: Figure a,b,c and d represent the unfolding pattern in a fiber with 15, 25, 50 and 100% of crosslinkers for System2(5×10.

Although our result is very much similar to fibrin gel shearing results, there is a lack of results for fibrin gel stretching, which shows a plateau-like region. To look for the reason, we have come up with the possible solution that the unfolding may not occur in so much abundance, as in a complex structure of a network, where there are other ways of relieving stress. Therefore, we introduce the case when only the stressed bond attached to the node gets unfolded, and net unfolding in a time step is 10% of eligible bonds. In the Figure 13, we use the condition that if the bond length is greater than 8 nm, we update the rest length to 6.25 nm. This plot shows a bilinear behaviour, surprising the unfolding region, which is as per with the behavior seen in experimental studies of the fibrin network.

## CONCLUSION

In this paper, we have proposed a coarse-grained model of fibrinogen, which includes the mechanism of unfolding of domains at the monomer level, to study the effect of unfolding on the mechanical properties of fibrin fiber and network. This unfolding-incorporated coarse-grained polymer (UCGP) model for fibrinogen provides an important structural characteristic that reflects on the stretching behaviour of fibrinogen through step-wise unfolding of $\gamma$-nodule. It provides a bridge between the fibrin monomeric properties and the fibrin network by mapping the unfolding of the domains. The minimal bead-spring model captures the unfolding through the update in spring constant and rest-length upon building up a threshold stress on a spring representing the $\gamma$ chain. Our model well exhibits the behaviour as seen in AFM experiment and all-atom simulation (16).

Our focus was exclusively on the development of a minimal coarse-grained bead-spring model, which would be just enough to produce the results of all-atom simulation and AFM experiment, keeping the scope open to extend it to higher-order structures like fiber and network. The multiscale modeling discusses the upscaling of molecular-level behavior to larger structures and how molecular interactions give rise to macroscopic properties. This can be achieved from our model by introducing 'knob-hole' interactions and cross-linking of fibrin represented as nodes and edges. We speculate that the extra length of fiber upon unfolding of domains, which we introduced in the UCGP model, may show plateau-like behavior in the elastic modulus. Our primary results of fibers and networks based on the current model show the expected behavior as seen in experiments. As our model includes all the important mechanisms that may affect fibrin stretching mechanics, viz, crosslinking, knob-hole interaction from structural aspects, and domain unfolding from dynamical behavior, this model can provide a complete understanding of the fibrin mechanics, giving an idea of the dominating mechanism in a given regime of stress.

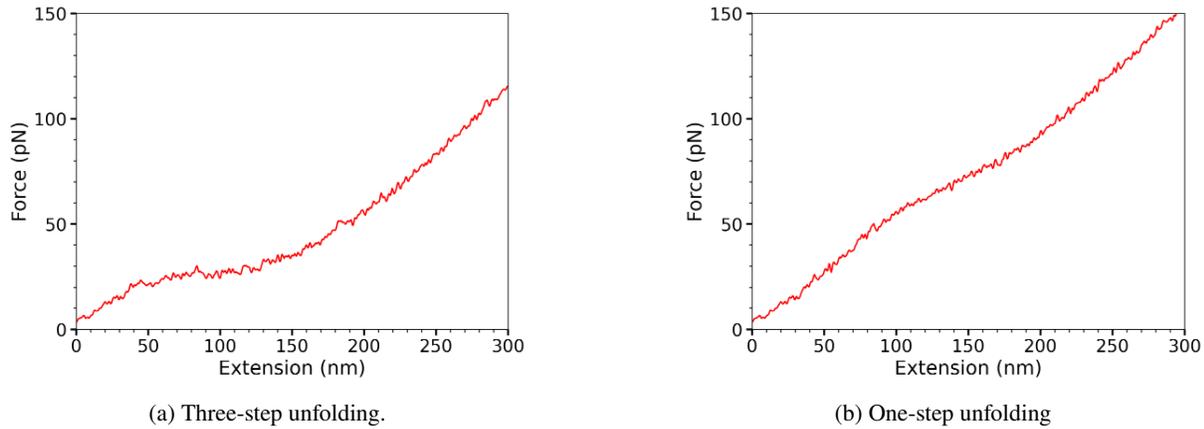

(a) Three-step unfolding.  (b) One-step unfolding

Figure 12: The force-extension behavior of the triangulated network a) with 3-step unfolding, and b) with 1-step unfolding.

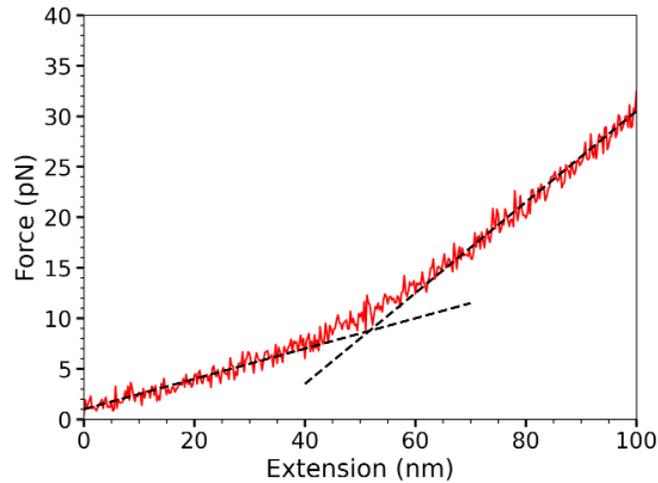

Figure 13: The force-extension behavior for a triangulated fibrin network with one-step and diluted unfolding.

This model is the first of its kind that incorporates the domain unfolding through a bead-spring model, which can be used to build multiscale structures. In our model, we change the stiffness of springs while keeping the ratio the same, therefore, we can control the effective fiber stiffness, which would then represent fibers of different thickness, keeping the qualitative behaviour unchanged. Additionally, this model can also be applied to polymers with sacrificial bonds, which show a similar force-extension curve (33, 34), by changing the update criteria of rest-length upon breaking the sacrificial bonds. Therefore, this model can be very useful in studying the mechanical properties of the bio-polymers having the protein unfolding or sacrificial bond lengths that greatly affect the mechanical properties.